\documentclass{article}

\usepackage{amsmath,amssymb,latexsym}
\usepackage[mathscr]{eucal}

\newcommand{\om}{\omega}

\newcommand{\ph}{\varphi}
\newcommand{\BR}{\mathbb{R}}
\newcommand{\Dl}{\Delta}
\newcommand{\Ga}{\Gamma}
\newcommand{\ga}{\gamma}
\newcommand{\dl}{\delta}
\newcommand{\1}{^{-1}}

\newcommand{\asl}{\text{\rm sl}}
\newcommand{\BS}{\mathbb{S}}
\newcommand{\BC}{\mathbb{C}}
\newcommand{\f}{\frac}
\newcommand{\ii}{\int\limits}
\newcommand{\ov}{\overline}
\newcommand{\wt}{\widetilde}

\newcommand{\PRL}{Phys. Rev. Lett. }

\newcommand{\CMP}{Commun. Math. Phys. }
\newcommand{\LMP}{Lett. Math. Phys. }

\begin{document}

\begin{center}
{\LARGE Discrete evolution for the zero-modes}\\
\smallskip
{\LARGE of the Quantum Liouville Model}\\
\smallskip

{L.~D.~Faddeev, A.~Yu.~Volkov}\\

{St.Petersburg Department \\
of Steklov Mathematical Institute}
\end{center}

\bigskip

\begin{center}
    Abstract
\end{center}

    The dynamical system for the zero modes of the Liouville Model,
    which is separated from the full dynamics for the discrete
    shifts of time
$ t \to t + \pi $,
    is investigated. The structure of the modular double in quantum case is
    introduced.

\section{Introduction}
The Liouville model, based on the famous Liouville equation
\cite{c1}
\begin{equation}
\label{L}
    \ph_{tt}-\ph_{xx}+e^{2\ph}=0
\end{equation}
for the real function $\ph(x,t)$ on the two-dimensional
space-time, plays an important role in mathematics and
physics. In quantum variant it gives an example of
Conformal Field Theory and enters as a basic ingredient
into Polyakov's theory of noncritical string 
\cite{c2}. In this
application the space-time is taken as a cylinder
$\BR^1\times\BS^1$, so that $\ph(x,t)$ satisfies the
periodicity condition
\begin{equation}
\label{per}
    \ph(x+2\pi,t)=\ph(x,t),
\end{equation}
and it is especially important to describe
the spectrum of primary fields or, in other words, zero
modes. This picture was thoroughly investigated in the
early 80-ties by canonical quantization 
\cite{c3}--\cite{c5} and more
recently by means of the theory of representations for the
quantum group $\mathcal{U}_q(\asl(2,\BR))$ in 
\cite{c6}, \cite{c7}. In this paper
we shall develope an alternative approach, based on the methods
of old papers
\cite{c5}, \cite{c8}. The main
result is the following observation: the zero modes can be
combined into dynamical variables $Z_n$ with simple
evolution equation, corresponding to the discrete shift
$t\to t+\pi$, which looks as follows
\begin{equation}
\label{sys}    
Z_{n+1}Z_{n-1}=1+Z^2_n
\end{equation}
(with quantum correction introduced in the main text). Thus
for such time shifts the zero modes completely decouple
from the infinite number of the oscillator degrees of
freedom. The main interest is in the spectrum of the
corresponding quantum Hamiltonian. We shall confirm the
known result, that there exists a quantization, for which
this spectrum is simple, continuous and positive. However
we believe, that our approach could lead to an alternative
quantization, which is now under investigation.

Let us mention in passing, that the system 
(\ref{sys}) gives a
basic example in the modern theory of cluster algebras
\cite{c9}.
The symplectic approah of our paper could be instructive
for the specialists in this topic.

The plan of the paper is as follows.

In section 1, we derive the system 
(\ref{sys}) in classical
setting.
In section 2 the zero curvature representation of the Liouville
model is reminded. Explicit solution of
(\ref{sys}) and its canonical
interpretation is given in section 3. Section
4 is devoted to quantization. Finally, in section 5, we
discuss a new feature, inherent only to quantum case -- the
modular dual dynamical model. It is in this point, where our
hope for an alternative quantization lyes.

The work on this project was parthly supported by the
RFBR, grant 05-01-00922 and programm ``Mathematical
problems of nonlinear dynamics'' of Russian Academy of Sciences.

\section{Derivation of (3)}

As is well known, the general solution of the Liouville
equation 
(\ref{L}) is given by Liouville formula
\cite{c1}
\begin{equation}
\label{Lf}
    e^{2\ph(x,t)}=-4\frac{f'(x-t)g'(x+t)}{(f(x-t)-g(x+t))^2},
\end{equation}
where $f(x)$ and $g(x)$ are arbitrary real functions of one
variable and $f',\,g'$ are their derivatives. For a given
$\ph(x,t)$ we shall call $f$ and $g$ its representative
functions. The correspondence 
$$
\ph\leftrightarrow(f,g)
$$ 
is not one to one. 
In particular $\ph$ is invariant
with respect to a fractional-linear (M\"obius) transformation
\begin{equation}
\label{Mtr}
    f\to\frac{af+c}{bf+d},\qquad g\to\frac{ag+c}{bg+d}
\end{equation}
for any real unimodular matrix $\begin{pmatrix} a&b\\ c&d
\end{pmatrix}$.

For the RHS of 
(\ref{Lf}) to be positive and finite it is enough
to require, that $f$ and $g$ are correspondingly
monotonously decreasing and increasing,
\begin{equation}
\label{posc}
    f'<0,\qquad g'>0.
\end{equation}
These conditions are compatible with transformaion 
(\ref{Mtr}). We
shall refer to 
(\ref{posc}) as the positivity condition.

The periodicity 
(\ref{per}) can be achieved if the representative
functions satisfy the relation
$$
f(x+2\pi)=\frac{Af(x)+C}{Bf(x)+D},\qquad
g(x+2\pi)=\frac{A g(x)+C}{Bg(x)+D}
$$
with the monodromy matrix
$$
T=\begin{pmatrix} A&B\\ C&D
\end{pmatrix},
$$
depending on the solution
$\ph(x,t)$ and choice of the representative functions $f$
and $g$. The trace of monodromy
$$
A+D=2\cosh P
$$
is representation invariant and define an important
dynamical variable $P$, which is called the quasimomentum
of zero modes.

With these preliminarilies we can turn to the derivation of
(\ref{sys}). Take a shift $\Dl$ -- positive number and construct
combination of the representative functions $f$ and $g$
which is a cross-ratio
$$
Y(x,t)[\ph]= -
\frac{(f(x-t+\Dl)-f(x-t-\Dl))(g(x+t+\Dl)-g(x+t-\Dl))}
{(f(x-t+\Dl)-g(x+t+\Dl))(f(x-t-\Dl)-g(x+t-\Dl))}.
$$
This combination is positive due to the positivity
condition. It follows from the M\"obius invariance, that
$Y[\ph]$ does not depend on the choice of the
representative functions $f$ and $g$. It is also clear,
that for $\Dl\to 0$
$$
Y(x,t)[\ph]\sim\Dl^2 e^{2\ph(x,t)}.
$$

Now consider four values of $Y(x,t)[\ph]$
\begin{align*}
Y_N=& Y(x,t+\Dl)[\ph], \quad Y_W=Y(x-\Dl,t)[\ph], \\
Y_E=& Y(x+\Dl,t)[\ph], \quad Y_S=Y(x,t-\Dl)[\ph].
\end{align*}
For fixed $x,t$ these four cross-ratios depend on six
numbers $f(x-t)$, $f(x-t\pm\Dl)$, $g(x+t)$, $g(x+t\pm\Dl)$
and due to M\"obius invariance only three of them are
independent. Therefore there exists a relation between them
and indeed it is easy to check that this relation has
the form
\begin{equation}
\label{Yrel}    
    Y_NY_S=\frac{Y_W\,Y_E}{(1+Y_W)(1+Y_E)}.
\end{equation}
This consideration was used before in
\cite{c10}, where the discretized version
of the Liouville equation was introduced. Mention also,
that 
(\ref{Yrel}) gives the simplest example of so called $Y$-systems
\cite{c11}.

Now recall that
$ Y(x,t)[\ph] $ is
$ 2\pi $-periodic in $ x $.
    Hence, if
$ \Dl = \pi $,
    then
$ Y_W = Y_E $
    and eq
(\ref{Yrel})
    becomes
$$
    Y_N Y_S = \frac{Y_W^2}{(1+Y_W)^2} = \frac{Y_E^2}{(1+Y_E)^2} ,
$$
    or in terms of inverse square roots
$ Z = 1/\sqrt{Y}$
$$
    Z_N Z_S = 1+Z_W^2 = 1+ Z_E^2 .
$$
    Hence, clearly, the sequence
$$
    Z_n = \frac{1}{\sqrt{Y((n+1)\pi,n\pi)}}
$$
    satisfies the recurrence relation
$$
    Z_{n+1} Z_{n-1} = 1 + Z_n^2 .
$$

\section{Zero curvature and the representative functions}
    We remind here the zero curvature of the Liouville model, introduced in
\cite{c5}.
    The pair of Lax operators
$$
L_1=\f d{dx}-L,\quad L_0=\f d{dt}-M
$$
with traceless $2\times 2$ matrices
\begin{align}
\label{mL}
L=&\f12 \begin{pmatrix} \ph_t(x,t) & e^{\ph(x,t)}\\
    e^{\ph(x,t)}& -\ph_t(x,t)\end{pmatrix} ; \\
\label{mM}
M=&\f12 \begin{pmatrix} \ph_x(x,t) & - e^{\ph(x,t)}\\
    e^{\ph(x,t)}& -\ph_x(x,t)\end{pmatrix} ,
\end{align}
    produce the equation 
(\ref{L})
    as a zero curvature condition
$$
L_t-M_x+[L,M]=0.
$$
Mention, that in contrast with more generic case (see,
e.g.,~\cite{c12}) this Lax pair does not contain spectral
parameter, so it corresponds to the finite dimensional Lie
algebra $\asl(2,\BR)$ rather than to its affine extension.
In fact, one can get these Lax operators as a degenerate
case of those for the Sine-Gordon equation from
\cite{c12}.

Let
$$
T(x,t)= \begin{pmatrix} A(x,t)&B(x,t)\\ C(x,t)&D(x,t)
\end{pmatrix} 
$$
be a solution of compatible equations
\begin{equation}
\label{ceq}
    T_x=LT,\qquad T_t=MT
\end{equation}
with initial condition
$$
T(0,0)=I.
$$
This solution is a unimodular hyperbolic matrix
$$
\det T=AD-BC=1;\qquad \text{\rm tr}\, T=A+D>2.
$$
    and satisfies the quasiperiodycity condition
\begin{equation}
\label{qcond}    
    T(x+2\pi,0) = T(x,0) T(2\pi,0)
\end{equation}
The representative functions $f(x)$, $g(x)$ are easily
expressed via matrix elements of $T(x,t)$. In more detail,
consider the ratios
\begin{equation}
\label{rat}
    f(x,t) = \frac{A(x,t)}{B(x,t)} , \quad
    g(x,t) = \frac{C(x,t)}{D(x,t)} .
\end{equation}
    It easily follows from
(\ref{ceq}),
    that
$$
    f_x = - f_t = - \frac{e^{\ph}}{2B^2} , \quad
    g_x = g_t = \frac{e^{\ph}}{2D^2} .
$$
    Hence 
$$
    f(x,t) = f(x-t) , \quad g(x,t) = g(x+t) .
$$
    Furthermore, we have from
(\ref{rat})
$$
    f - g = \frac{AD-BC}{BD} = \frac{1}{BD}
$$
    so that
$$
    -4 \frac{f'(x-t)g'(x+t)}{(f(x-t) - g(x-t))^2} = e^{2\ph} .
$$
    Finally the quasiperiodicity
(\ref{qcond})
    in terms of
$ f $ and $ g $
    looks as follows
\begin{equation}
\label{fgc}    
    f(x+2\pi) = \frac{Af(x)+C}{Bf(x)+D} , \quad
    g(x+2\pi) = \frac{Ag(x)+C}{Bg(x)+D} , 
\end{equation}
    where
$ A,B,C,D $
    are matrix elements of
$$
    T = T(2\pi,0) =
\begin{pmatrix}
    A & B \\
    C & D
\end{pmatrix} ,
$$
    so that
$ T(2\pi, 0) $
    indeed plays the role of monodromy. Thus
$ f(x) $ and $ g(x) $
    give us a particular set of the representative functions,
    normalized by condition
\begin{equation}
\label{nc}    
    f(0) = \infty , \quad g(0) = 0.
\end{equation}

    Now we shall connect the initial data
$ Z_{-1}, Z_{0} $
    for our system for
$ Z_n $
    with the elements of monodromy.
    We have
\begin{align}
\label{Z1}
    Z_{-1} = & \bigl(Y(0,-\pi)\bigr)^{-1/2} 
	= \bigl(- \frac{(f(2\pi)-f(0))(g(0)-g(-2\pi))}{
	    (f(2\pi)-g(0))(f(0)-g(-2\pi))}\bigr)^{-1/2} = \\
\nonumber
	=& \bigl(-\frac{g(-2\pi)}{f(2\pi)}\bigr)^{-1/2}
	    = \frac{A}{\sqrt{BC}} ; \\
\label{Z0}
    Z_{0} = & \bigl(Y(\pi,0)\bigr)^{-1/2} 
	= \bigl(- \frac{(f(2\pi)-f(0))(g(2\pi)-g(0))}{
	    (f(2\pi)-g(2\pi))(f(0)-g(0))}\bigr)^{-1/2} = \\
\nonumber
	=& \bigl(\frac{g(2\pi)}{f(2\pi)-g(2\pi)}\bigr)^{-1/2}
	    = \frac{1}{\sqrt{BC}} .
\end{align}
    We used here the normalization
(\ref{nc})
    and quasiperiodicity
(\ref{qcond}).

    We finish this section by expressing the monodromy via
$ Z_{-1}, Z_{0} $.
    From
(\ref{Z1}), (\ref{Z0}) we have
\begin{equation*}
    A = \frac{Z_{-1}}{Z_{0}} , \quad BC = \frac{1}{Z_{o}^{2}} .
\end{equation*}
    Then
\begin{equation*}
    D = \frac{1+BC}{A} = \frac{1+ Z_{0}^{-2}}{Z_{-1}}
	= \frac{1+Z_{0}^{2}}{Z_{-1}Z_{0}} 
	= \frac{Z_{0}}{Z_{1}} + \frac{1}{Z_{-1}Z_{0}} .
\end{equation*}
    The only unknown remains the ratio
$ \alpha = B/C $.
    In terms of
$ Z_{-1}, Z_{0} $ and $ \alpha $
    the monodromy looks as follows
\begin{equation*}
    T = \begin{pmatrix}
	\frac{Z_{-1}}{Z_{0}} & \frac{\alpha}{Z_{0}} \\
	\frac{1}{\alpha Z_{0}} & \frac{Z_{0}}{Z_{-1}} + \frac{1}{Z_{0}Z_{-1}}
    \end{pmatrix} .
\end{equation*}
    
\section{Explicit solution and its canonical interpretation}
    It can be easily shown, that the solution of our system for
$ Z_{n} $
    can be solved rather explicitely via parametrization
\begin{equation}
\label{Zpar}
    Z_{n} = \frac{\cosh (Q+nP)}{|\sinh P|} ,
\end{equation}
    where
$ Q $ and $ P $
    are to be found from the initial data
\begin{equation*}
    Z_{-1} = \frac{\cosh (Q-P)}{|\sinh P|} , \quad
    Z_{0} = \frac{\cosh Q}{|\sinh P|} .
\end{equation*}
    Thus the evolution in new variables consists in the mere shift
\begin{equation*}
    P \to P , \quad Q \to Q+P .
\end{equation*}
    Variable
$ P $
    is conected with the invariant of the monodromy
\begin{equation*}
    A+D = \frac{1}{Z_{-1} Z_{0}}+ \frac{Z_{0}}{Z_{-1}} + \frac{Z_{-1}}{Z_{0}}
	= 2 \cosh P .
\end{equation*}
    Expression for
$ Q $
    is less transparent and we shall not present it here.
    Its connection with the fixed points of monodromy is discussed in
\cite{c5}.

    Let us give the canonical interpretation for these variables, returning
    first to the full Liouville model and following
\cite{c3}, \cite{c5}.
    Equation
(\ref{L})
    has a natural canonical presentation with the phase space coordinates
    being two functions on the circle, namely
\begin{equation*}
    \ph(x) = \ph(x,0) , \quad \pi(x) = \ph_{t}(x,0)
\end{equation*}
    with Poisson brackets
\begin{equation}
\label{Pbr}
    \{\pi(x), \ph(x)\} = \gamma \delta(x-y)
\end{equation}
    and Hamiltonian
\begin{equation*}
    H = \frac{1}{2\gamma} \int_{0}^{2\pi} \bigl(\pi^{2}(x) 
	+ \ph_{x}^{2}(x) + e^{2\ph(x)}\bigr) dx .
\end{equation*}
    The positive coupling constant
$ \gamma $
    does not enter the equation of motion, however we shall retain
    it for future quantization.

    It follows from
(\ref{Pbr}),
    that the elements of the Lax matrix
$ L $ 
(\ref{mL})
    have the Poisson relations, which can be written as
\begin{equation}
\label{Prel}
    \{L(x) \stackrel{\otimes}{,} L(y)\} = [r , L(x)\otimes I+ I\otimes L(x)]
	\delta(x-y) ,
\end{equation}
    if we use the general notations from
\cite{c12}.
    Here
$ r $
    is a classical
$ r $-matrix (in fact historically the first example of such matrices)
\begin{equation*}
    r = \frac{\gamma}{2} 
    \begin{pmatrix}
	0 & 0 & 0 & 0 \\
	0 & -1 & 2 & 0 \\
	0 & 0 & -1 & 0 \\
	0 & 0 & 0 & 0
    \end{pmatrix} .
\end{equation*}
    Local relation
(\ref{Prel})
    leads to the relation for the monodromy
\begin{equation*}
    \{T \stackrel{\otimes}{,} T \} = [r , T\otimes T]
\end{equation*}
    or explicitely
\begin{align*}
    \{A, B\} &= \frac{\gamma}{2} AB , \quad \{A, C\} = \frac{\gamma}{2} AC \\ 
    \{D, B\} &= -\frac{\gamma}{2} BD ,\quad \{D, C\} = -\frac{\gamma}{2} CD \\ 
    \{A, D\} &= 2\gamma BC , \quad \{B, C\} = 0 \\ .
\end{align*}
    This allows to calculate brackets for the initial data using the
    expressions
(\ref{Z1}) and (\ref{Z0}).
    We get
\begin{equation*}
    \{Z_{-1}, Z_{0}\} = -\frac{\gamma}{2} Z_{-1}Z_{0} .
\end{equation*}
    Finally it follows from relation
\begin{equation*}
    d \ln Z_{-1} \wedge d \ln Z_{0} = - dP \wedge dQ
\end{equation*}
    that $ P $ and $ Q $ have canonical brackets
\begin{equation*}
    \{P, Q\} = \frac{\gamma}{2} .
\end{equation*}

    However the phase space for $ P $ , $ Q $ is not the full plane,
    but rather half of it. Indeed the map
\begin{equation*}
    (P,Q) \to (Z_{-1}, Z_{0})
\end{equation*}
    for positive
$ Z_{-1}, Z_{0} $
    has as a preimage one of the half-planes
$ P >0 $ or
$ P <0 $.
    Thus the phase space of
$ P,Q $ is
$ \BR^{2}/Z_{2} $.
    It is clearly inconvenient for the quantization.

    The way out consists in the following: consider the parametrization
\begin{equation}
\label{Znpar}
    Z_{n} = e^{Q+nP} + \frac{1}{4 \sinh 2P} e^{-Q-nP} ,
\end{equation}
    which differs from
(\ref{Zpar})
    by the shift of
$ Q $
\begin{equation*}
    Q \to Q + \ln 2|\sinh P| ,
\end{equation*}
    which do not change the canonical structure.
    It is easy to see, that such
$ Z_{n} $
    are invariant under the action of canonical
    transformation~$K$ 
$$
K:P\to-P;\qquad Q\to-Q-\ln(4\sinh^2 P),
$$
which is a superposition of a simple reflection
$$
K_0:P\to-P;\qquad Q\to-Q
$$
and point transformation
$$
S:P\to P;\qquad Q\to Q-\ln(4\sinh^2P).
$$
The latter is defined via action
$$
Q\to Q-\f{dS(P)}{dP}
$$
with
\begin{equation}
\label{S}
    S(P)=\ii^P\ln( 4 \sinh^2 P) dP.
\end{equation}
The last integral reduces to dilogarithm
$$
\ii\ln\Big(t-\f1t\Big)^2\f{dt}t=2\ii(\ln(t-1)+\ln(t+1)-\ln
t)\f{dt}t.
$$
    It is clear, that $ K $ is a reflection
\begin{equation*}
    K \circ K = id
\end{equation*}
    which interchanges the half-planes.
    Its connection with the Weyl reflection for the monodromy is evident.

    All this allows us to consider the variables
$ Z_{n} $
    on the whole plane
$ \BR^{2} $
    and approach to quantization correspondingly, taking into account
    their invariance under reflection.

\section{Quantization}
    Formula 
(\ref{Znpar})
    gives $Z_n$ as rational function of exponents
$e^P$ and $e^Q$ with relation
$$
\{e^P,e^Q\}=\ga e^Pe^Q.
$$
Under canonical quantization
$$
\{P,Q\}\to\f i\hbar[P,Q]
$$
these exponents turn into the Weyl operators $u$, $v$ with
commutation relation
$$
uv=q\1vu,\qquad q=e^{i\ga/2} 
$$
(we put $\hbar=1$).
    Naive ordering of factors in
(\ref{Znpar}) leads to the formula
\begin{equation}
\label{fZ}
Z_n=q^{-n/2}vu^n+q^{n/2}\f1{u-u\1} u^{-n}v^{-1} \f1{u-u\1}.
\end{equation}
    It is easy to check, that
$$
    Z_{n-1} Z_{n} = q Z_{n} Z_{n-1}
$$
and
$$
Z_{n+1}Z_{n-1}=1+q\1Z^2_n.
$$
Thus formula 
(\ref{fZ})
    defines a quantization of the system 
(\ref{sys}).
If operators $u$, $v$ are self-adjoint and positive, the
same is true for $Z_n$, if
$ \bar{q} = q^{-1} $,
    which is valid for real
$ \ga $.

Unitary operator $U$, such that
$$
U\1 u U=u;\qquad U\1 v U=q^{-1/2} vu ,
$$
realized also the evolution
\begin{equation}
\label{evol}    
    Z_{n+1}=U\1 Z_n U.
\end{equation}
The Hamiltonian $H$ is given by
\begin{equation}
\label{Uham}    
U=e^{-2\pi i H} .
\end{equation}

The Weyl operators $u$ and $v$ usually are realized via
multiplication and shift. We shall use the explicite
formulas with a particular parametrization of the coupling
constant $\ga$. Its convenience will become clear later.
First we renormalize $\ga$ putting
$$
\ga= 2\pi \tau ,
$$
so that the phase factor $q=e^{i\ga/2}$ becomes $q=e^{i\pi \tau}$,
reminiscenting an object of the automorphic functions theory.
Following the tradition, stemming from Weierstrass, we
introduce two complex parameters $\omega$, $\omega'$ with normalization
$$
\omega\omega'=-\f14
$$
and put
$$
\tau=\f{\omega'}\omega.
$$
For $\tau$ to be real and positive, we take $\om,\om'$ as pure
imaginary with positive imaginary part
$$
\ov \om=-\om,\qquad  \ov \om'=-\om'.
$$
Operators $u$ and $v$ act on the Hilbert space $L_2(\BR)$
with elements $\psi(s)$, $-<s<\infty$, and scalar product
$$
(\psi',\psi)=\ii^\infty_{-\infty}\ov{\psi}'(s)\psi(s)ds
$$
as follows
\begin{equation}
\label{Wsys}
    u\psi(s)=e^{i\pi s/\om}\psi(s),\qquad v\psi(s)=\psi(s+\om').
\end{equation}
These operators are unbounded, an admissible domain of
definition $D$ consists of function $\psi(s)$, analitic in
the whole complex plane $\BC$ and rapidely vanishing along
contours, paralled to the real axis. Functions of the form
$$
\psi(s)=e^{-s^2}P(s),
$$
where $P(s)$ -- polinomial, are representative for $D$.
With this definition operators $u$ and $v$ are essentialy
self-adjoint and positive definite. The same is true for~$Z_n$.

The Hilbert space $L_2(\BR)$ realizes the quantization of
the phase space, which is the whole plane $\BR^2$. The reduction
to the half-plane after quantization should use the quantum
analogue of the canonical transformation $K$. Let us
construct the corresponding unitary operator such that
$$
KZ_n=Z_nK,\qquad K^2=I.
$$
As in classical case we put
$$
K=K_0S,
$$
where $K_0$ is a simple reflection
$$
K\1_0uK_0=u\1,\qquad K\1_0vK_0=v\1
$$
and $S$ realizes the transformatation
\begin{equation}
\label{Str}    
S\1uS=u;\qquad S\1vS=(u-u\1)v(u-u\1) .
\end{equation}
Due to commutativity with $u$, $S$ is a multiplication operator
$$
S\psi(s)=S(s)\psi(s).
$$
From quasiclassical consideration $S$ must be expressed via
exponent of the deformed classical action from 
(\ref{S}), 
    which involves dilogarithm. Candidate for this is already known,
see 
\cite{c13}, \cite{c14}. It is a meromorphic function $\ga(\xi)$
defined via functional equations
\begin{equation}
\label{feq}
\f{\ga(\xi+\om')}{\ga(\xi-\om')}=1+e^{-i\pi\xi/\om} ,
\end{equation}
having properties
\begin{equation}
\label{eprop}
\ga(\xi)\ga(-\xi)=e^{i\pi\xi^2},\qquad \ov{\ga(\xi)}=\f1{\ga(\ov\xi)} .
\end{equation}
The proper definition and beautiful properties of this
function, which generalizes both exponent, dilogarithm and
$\Ga$-function, are described in
\cite{c15}.

Let $M$ be a multiplication operator by function
$$
M(s)=e^{-2\pi is(s+\om'')}\ga(2s+\om'') ,
$$
where
$$
\om''=\om+\om'.
$$
It follows from 
(\ref{eprop}), that for real $s$
$$
M(-s)=\ov M(s)
$$
and we get from 
(\ref{feq})
$$
\f{M(s+\om')}{M(s)}=i\big(q e^{\f{i\pi s}w}-q\1e^{-\f{i\pi s}w}\big)
$$
Thus
$$
M\1vM=\f{M(s+\om')}{M(s)}v=i(qu-q\1u\1)v=iv(u-u\1).
$$
Now we put
$$
S(s)=\f{M(s)}{M(-s)}
$$
    and finally get
$$
S\1vS=(u-u\1)v(u-u\1)
$$
as was required in 
(\ref{Str}). Operator $S$ is unitary
$$
\ov{S(s)}=\f{\ov{M(s)}}{\ov{M(-s)}}=\f{M(-s)}{M(s)}=S\1(s)=S(-s)
$$
so that
$$
K^2=(K_0S)^2=K_0SK_0S=S\1S=I;
$$
thus $K$ is indeed a reflection.

We can introduce a pair of projectors
$$
\Pi_\pm=\f12(I\pm K) ,
$$
which commute with the operators $Z_n$. Thus each of the reduced
Hilbert spaces
$$
\mathcal{H}_\pm=\Pi_\pm L_2(\BR)
$$
can be considered as a physical Hilbet space in which
operators $Z_n$ are defined. This reduction is quantum
analogue of using half-plane as a classical phase space.

    The evolution 
(\ref{evol})
    is given by
$$
U\psi(s)=e^{-2\pi iH}\psi(s)=e^{-2\pi is^2}\psi(s),
$$
in other words, $H$ is operator of multiplication by $s^2$.
Its spectrum is continuous and positive. In the whole
$L_2(\BR)$ it has multiplicity 2 with generalized
eigenfunctions $\dl(s+k)$ and $\dl(s-k)$. However in the
reduced space only one combination is an eigenfunction; for
example in the space $\mathcal{H}_+$ it is
$$
\dl(s-a)+S(k)\dl(s+a).
$$
So the physical spectrum of $H$ is simple. We can not help
mentioning the analogy of the last formula with those of the scattering
theory involving the  incoming and outgoing waves, connected by
$S$-matrix. In fact, this resemblance is not accidental. We
hope to return to it, discussing a natural quantum
deformation of the Gindikin--Karpelevich fornula in the future.

\section{Modular double}
As was already mentioned some time before 
\cite{c16},
    the Weyl system 
(\ref{Wsys}),
    acting in $L_2(\BR)$, has a dual partner
$$
\wt u\psi(s)=e^{i\pi s/\om'}\psi(s),\qquad \wt v\psi(s)=\psi(s+\om)
$$
with relations
$$
\wt u \wt v=\wt q\1\wt v\wt u,\qquad \wt q=e^{i\pi/\tau}.
$$
It is here where we see the convenience of our variables
$\om,\om'$: dualization corresponds to the
interchange~$\om\leftrightarrow \om'$.

Operators $\wt u$, $\wt v$ are also positive self-adjoint
for real positive $\tau$. They have simple commutation
relation with $u$, $v$
$$
u\wt u=\wt u u,\quad v\wt v=\wt v v;\qquad u\wt v=-\wt v
u,\quad v\wt u=-\wt u v.
$$
    We can use them to construct dual variables $\widetilde Z_n$ by
    formula 
(\ref{fZ})
    with substitution of $u$, $v$ by dual ones
$\wt u$, $\wt v$. It is clear, that operator $Z_n$ and $\wt
Z_n$ commute. The operator $K$ and Hamitonian $H$ are
invariant under interchange $\om\leftrightarrow \om'$ and so
play for $\wt Z_n$ the same role as for~$Z_n$.

Thus we observe, that after quantization we acquire a second
dynamical system acting in the same Hilbert space. 
For positive coupling constant the second
system is independent of the first, both of them are
selfconsistent. However our construction allows us to use
another regime for $\tau$, which is needed for the
application of Liouville Model to noncritical string.
Indeed, as is well known,
the central charge $C$ of
the quantum Liouville Model is given by the formula
$$
C=1+\Big(\tau+\f1\tau+2\Big)
$$
and it is a real number for real $\tau+\f1\tau$, so that in
addition to positive $\tau$ we can use $\tau$ such that
\begin{equation}
\label{tprop}
\ov \tau=\f1\tau,\qquad |\tau|=1.
\end{equation}
For real positive $\tau$ we have $C>25$, but for $\tau$ from
(\ref{tprop})
$$
1<C<25.
$$
It is exactly this regime, which is relevant to
noncritical string and we can use our doubling to treat
this case.

Of course, the complex coupling constant raises question of
self-adjointness and unitary. However, in regime
(\ref{tprop}),
which is described by condition
$$
\ov \om=-\om',
$$
we have the involution
\begin{equation}
\label{inv}    
\wt Z^*_n=Z_n,
\end{equation}
which interchanges our commuting systems. Abstractly this
involution is similar to
$$
(a\otimes b)^*=b^*\otimes a^*
$$
in the tensor products. Thus we can consider 
(\ref{inv}) as a
legitimate variant of the reality condition, valid in the
regime 
(\ref{tprop}). The Hamiltonian 
(\ref{Uham}) can be used here as well,
so it is the same as for the case of positive~$\tau$.

This apparently supports the conviction of the papers
\cite{c6}, \cite{c7}, \cite{c17}
about smooth continuation of characteristics of Liouville
Model, such as $N$-point correlation functions. calculated
for $C>25$, through the threshold $C=25$. However we still
hope, that one can find an alternative quantization of
$Z_n$, $\wt Z_n$ in the regime
(\ref{tprop}), inequivalent to the
described here. This could vindicate our approach as not
purely methodological.

One reason for optimism is the recent result of one of the
authors (LDF) about discrete series of representation for
algebra $\mathcal{U}_q(\asl(2))$
\cite{c18} in regime
(\ref{tprop}). 
Work in this direction will be
continued. 

\section{Conclusions}
In this paper, we separated a finite-dimensional dynamical
system for the zero modes of the Liouville model in its
classical and quantized formulation. The new feature is the
modular duality in quantum case suitable to treat the case
of strong coupling. Let us add one more statement.

In \cite{c10}, the lattice variant of the Liouville Model was
proposed. The considerations of this paper could be
straightforwardly applied to this case also. The equations
for the zero modes are exactly the same as in this paper. We
belive, that this is another example of separation of
degrees of freedom in CFT both in continuous and lattice
variants with zero modes not depending on lattice
regularization. In less explicite form this feature was
already observed in 
\cite{c19} for the Wess--Zumino Model.

\end{document}